\documentclass[twocolumn]{aastex631}

\newcommand{\hb}{\hbox{H$\beta$}}

\newcommand{\gsim}{\lower.5ex\hbox{$\; \buildrel > \over \sim \;$}}
\newcommand{\lsim}{\lower.5ex\hbox{$\; \buildrel < \over \sim \;$}}

\newcommand{\oii}{\hbox{[O\,{\sc ii}]}}
\newcommand{\oiii}{\hbox{[O\,{\sc iii}]}}

\begin{document}


\title{Serendipitous Catch of a Giant Jellyfish: an Ionized Nebula around 3C 275.1 with 170 kpc Long Tails}

\correspondingauthor{Junfeng Wang}
\email{jfwang@xmu.edu.cn}

\author{Qinyuan Zhao}
\affiliation{Department of Astronomy, Xiamen University, Xiamen, Fujian 361005, China}

\author[0000-0003-4874-0369]{Junfeng Wang}
\affiliation{Department of Astronomy, Xiamen University, Xiamen, Fujian 361005, China}

\author[0000-0002-8237-0237]{Zhenzhen Li}
\affiliation{Key Laboratory for Research in Galaxies and Cosmology, Shanghai Astronomical Observatory, Chinese Academy of Sciences, 80 Nandan Road, Shanghai 200030, China}
\affiliation{Key Laboratory of Polar Science, MNR, Polar Research Institute of China, Shanghai 200136, China}



\begin{abstract}

3C 275.1 is a blue quasar at $z=0.55522$, hosting powerful outflows and residing in a complex environment. We present a serendipitously detected giant nebula surrounding 3C 275.1, which shows morphological features resembling those of objects known as ``jellyfish galaxy", with extremely long tails of ionized gas extending to 170 kpc in projection. We analyze its optical spectra taken by the MUSE on the VLT. The brighter part of this giant nebula exceeds 100 kpc, whose rotation curve does not flatten out, is very different from those of normal spiral galaxies. This system shares some characteristics common to those formed via ram pressure stripping (RPS), yet its long narrow tails and higher ionization are unusual compared to known tails in jellyfish galaxies, not fully consistent with a simple RPS scenario. Our photoionization simulation and the inferred short recombination timescale both suggest that besides the quasar 3C 275.1, an extra source of ionization is necessary to keep the gas ionized at such distance from the nucleus, which could be related to RPS, tidal interaction or AGN outflow, providing new evidence of active dynamical interaction of a galaxy with the intracluster medium.

\end{abstract}


\keywords{Galaxy evolution (594); Intracluster medium (858); Galaxy clusters (584); Ram pressure stripped tails (2126)}


\section{Introduction} \label{sec:intro}
\label{sec:introduction}

Environmental effects play a key role in galaxy evolution, specifically the transformation from late-type to early-type galaxies. Galaxies evolve primarily through mergers \citep{Toomre1972,Perry2023}, tidal interactions \citep{Mayer2006,Chung2007}, ram-pressure stripping (RPS; \citealt{Gunn1972,Luo2023}), viscous stripping \citep{Nulsen1982,Quilis2000}, galaxy harassment \citep{Moore1996}, and starvation \citep{Larson1980,Peng2015}. These interactions can remove the gas to rapidly quench star formation \citep{Roberts2019,Ciocan2020} or increase interstellar medium (ISM) density to enhance star formation \citep{Roberts2020,Cramer2021}.

Among these mechanisms, RPS has been known as the most efficient mechanism of gas removal from galaxies in cluster environments \citep{Dressler1980}. When a galaxy falls into a cluster through the intergalactic space filled with hot and diffuse gas, ram pressure from intra-cluster medium (ICM) gas applies a force to the gas component in the galaxy that can potentially unbind its gas reservoir. This process, which results in disturbed galaxy morphologies and trailing tails of stripped gas, is referred to as RPS (see \citealt{Boselli2022} for a review). \citet{Bekki2009} named these objects ``jellyfish galaxies" due to the morphologies of their tails. Observations and simulations both demonstrate important effects of RPS on galaxy evolution such as disk truncation, star formation quenching, central bulge build-up, formation of flocculent arms, the transformation of dwarf galaxies, and long filamentary structures (\citealt{Okamoto2003,Tonnesen2012,Ruszkowski2014,Kenney2014}).

In the past several years, more signatures of RPS have been observed in molecular gas (e.g. \citealt{Sivanandam2010,Sivanandam2014,Moretti2018,Moretti2020}), and ionized gas in the optical (e.g. \citealt{Fumagalli2014,Fossati2016,Bellhouse2019,Cramer2019,Azevedo2023}) and X-ray bands (e.g. \citealt{Sun2006,Sun2010,Poggianti2019}).  
JO201 with 94 kpc long tails which is falling into Abell 85, undergoes both RPS and AGN feedback at low redshift \citep{Bellhouse2019,George2019}. \citet{Boselli2019} detected two  massive star forming galaxies with tails that extend to 100 kpc at $z=0.7$. Recent studies further reveal that RPS can also help compressing the gas (e.g., \citealt{Roberts2020,TroncosoIribarren2020}), which may briefly enhance star formation during the early stages, and eventually remove the full ISM from the galaxy, leading to the quenching of star formation \citep{Vollmer2001,Tonnesen2007}.


It is critical to observe jellyfish galaxies at high physical resolution to better understand star formation activities at small-scale. Integral field spectroscopy (IFS) observations provide both spatial and spectral information, therefore it is very advantageous to investigate various physical properties of jellyfish galaxies with IFS observations. However, most IFS studies have been limited to the local universe, whereas jellyfish galaxies in massive clusters are mostly at intermediate redshift (e.g. \citealt{Kalita2019}). Hence case study of their local analogues can provide more insights.

In this paper, we present VLT/MUSE IFS observations of quasar 3C 275.1 at $z \sim 0.55$ and discovery of a giant nebula surrounding it, aiming to characterize its intergalactic environment and ISM. This system was serendipitouly identified during our study on galaxy scale quasar outflow using IFS archival data \citep{Zhao2023}. We carry out emission line analysis of the ionized gas in this newly identified jellyfish galaxy, which includes \oii, \hb\ and \oiii.
This paper is structured as follows: we describe observations and data reduction in
Section 2. In Sections 3 and 4, we measure the group environment and the nebula. In Section 5, we discuss the possible origin of the nebula.  
We conclude with a summary in Section 6. Throughout this paper, we adopt a cosmology with $H_{0}$=70 km s$^{-1}$ Mpc$^{-1}$, $\Omega_{m} = 0.3$, $\Omega_{\Lambda} = 0.7$.

\section{Observation and Data Reduction} 
\label{sec:section_name}
3C 275.1 was observed by VLT/MUSE in 2021 February under ESO program ID 106.2142.001, i.e., 0106.B-0564(A) (PI: Balmaverde). Results on the optical spectral properties of the nuclei of some radio-loud AGNs have been published \citep{Capetti2023}. The spectra were taken in the optical band (wavelength coverage $\lambda$ $\sim$ 4750 $\mathrm{\AA}$ - 9350 $\mathrm{\AA}$ in the observer's frame), covering 3054 $\mathrm{\AA}$ to 6012 $\mathrm{\AA}$ in the rest frame for 3C 275.1 at $z \sim 0.555$. The field of view (FOV) covers an average of $1' \times 1'$, and medium spectral resolution of R = 3500. The seeing varied during the observations from 0.65 arcsec to 1.2 arcsec, and the full width at half maximum (FWHM) of the point-spread function (PSF) in the averaged data set was $0.9\arcsec$. The total on-source integration time is 4 $\times$ 562 s.

After removing  cosmic rays from the raw data using the L.A. Cosmic procedure \citep{Dokkum2001}, we reduce the raw data using the ESO-MUSE pipeline. The final data cubes have a spatial scale of $0.2\arcsec \times 0.2\arcsec$. 
The estimated angular resolution is $\sim$ 0.9$\arcsec$ based on the FWHM of the PSF map, which roughly corresponds to 5.8 kpc at the redshift of $z = 0.55522$. We estimated the PSF for this observation using surface brightness profiles of broad emission lines (BELs), i.e., using a 2D Gaussian to fit the BELs map. 

\section{The Quasar 3C 275.1 and Its Environment} \label{sec:dr}

\subsection{Properties of 3C 275.1}  \label{sec:dr}

3C 275.1 with a central black hole of 10$^{8.3}$ M$_\odot$ \citep{McLure2006} at $z = 0.55522 \pm 0.00004$ \citep{AdelmanMcCarthy2008} is a FR II radio quasar.
The radio morphology of 3C 275.1 is a dogleg structure, and the two sides display an asymmetry in the Faraday depolarization \citep{Garrington1991}.
 Several authors have argued that the source lies in a rich cluster which could account for its distorted shape \citep{Stocke1985,Ellingson1991,Hardcastle1999,Crawford2003}, \citet{Liu1990} note that there is a galaxy near the southern component which may be interacting with the southern lobe.

3C 275.1 is hosted by an elliptical galaxy although the colors are somewhat bluer than normal \citep{Hintzen1986a}, and is surrounded by a number of faint galaxies \citep{Hintzen1981}. \citet{Ellingson1994} measured 35 of these galaxies and only found five to be within about 300 km s$^{-1}$ of the quasar redshift. 
Previous work suggests that 3C 275.1 is located at the centre of the gravitational potential well of the group \citep{KrempecKrygier1998}.  An extremely large elliptical nebula whose major axis roughly perpendicular to the radio axis (e.g. \citealt{Stocke1985}) and exceeding 100 kpc (14\arcsec) underlying 3C 275.1 \citep{Hintzen1983}, shows a ``solid-body rotation curve" extending 40 kpc from the quasar nucleus \citep{Hintzen1986}. The nebula may be relic from multiple tidal interactions with cluster/group members, or it may be the cooling flow centered on the quasar. In the latter case, the quasar was considered to be a possible ``proto-cD" galaxy \citep{Hintzen1986}. In this IFS observation, we also identified a large nebula around 3C 275.1, confirming the results of \citet{Hintzen1983} and \citet{Hintzen1986}. Moreover, the nebula we detected is much larger than previous results, extending to $\sim$170 kpc. \citet{Crawford2003} reported on the spatially extended soft X-ray emission surrounding 3C 275.1. The X-ray emission is most likely thermal emission from the ICM of a cluster, or possibly associated with the radio plasma \citep{Crawford2003, Belsole2006}.
We overlays the X-ray, radio, and optical nebula contours (Figure~\ref{fig:fig2}e) and find that the extended X-ray emission is preferentially aligned along the radio jet direction, while the optical filamentary nebula is located between radio/x-ray jet hotspots and is roughly perpendicular to jet axis.

\subsection{Group Membership}  \label{sec:dr}

\begin{figure}[htb]
\centering
\includegraphics[width=0.47\textwidth]{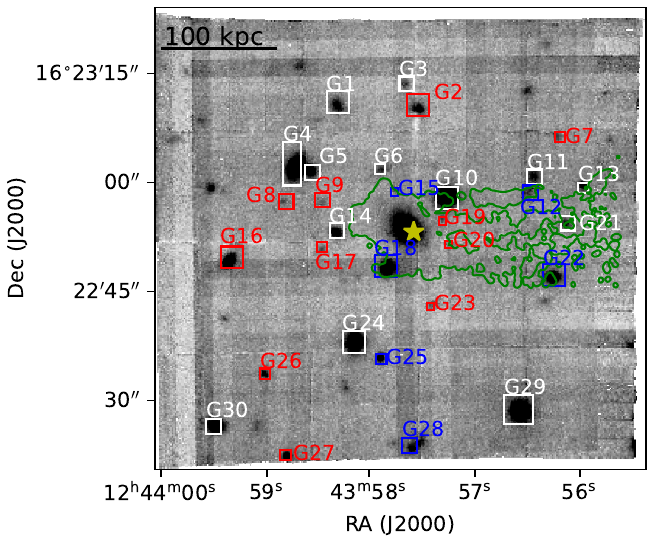}
\caption{MUSE white-light image in linear scale over the full field of view of the observations. The five group members that are unambiguously identified are marked
with blue rectangles, while the foreground or background sources are marked
with white rectangles, and sources with no identified redshifts are marked
with red rectangles. The yellow star marks the position of the quasar 3C 275.1. The morphology of \oiii\ is marked with green contours. The physical scale at the redshift of the group is indicated at the top left.
\label{fig:fig1}}
\end{figure}

In order to determine which galaxies belong to the ``group" and which are foreground or background sources, we have estimated spectroscopic redshifts of the sources in the region covered by
the ionized gas using their emission or absorption spectral features. Using available MUSE data (Figure~\ref{fig:fig1}). Thirty continuum sources are identified, and five group members are unambiguously spectroscopically identified with redshifts between $0.5535 \leq z \leq 0.5595$ in the FOV. Figure~\ref{fig:fig1} shows an image from the MUSE observations of the field where we mark the quasar and surrounding galaxies with their IDs, together with the morphology of nebula. Blue, white and red rectangles represent that the galaxy is a group member, a foreground or background source, and uncertain source, respectively.  The IDs, galaxy coordinates, photometric redshifts from SDSS Data Release 12 \citep{Alam2015a}, spectroscopic redshifts  and $\Delta v$ with 3C 275.1 are reported in Table~\ref{tab:tab1}. There could be some dwarf galaxies with weak continuum that have not been identified. The velocity range of these 5 group members is $\sim$ 1200 km s$^{-1}$. The 4 closest galaxies have a velocity range of only 650 km s$^{-1}$, possibly consistent with being members of a group. Such complex environment could result in galaxy interactions that contribute to redistribution of gas to large scales (e.g. \citealt{Johnson2024}).

\begin{table}[htb]\footnotesize
\caption{Summary of Galaxies in the Field of 3C 275.1.}
\hspace{-1in}
\begin{tabular}{cccccc}
\hline
\hline
 ID & RA & Dec &  photo $z$  & spec $z$ & $\Delta v$\\
\hline
G1    & 12:43:58.31 & +16:23:10.27 & 0.409 $\pm$ 0.1071 &  0.2367 &\\
G2    & 12:43:57.55 & +16:23:09.78 & 0.511 $\pm$ 0.0626 &  &\\
G3     & 12:43:57.65 & +16:23:13.38 & - & 0.4909 & \\
G4   & 12:43:58.71 & +16:23:01.87 & 0.241 $\pm$ 0.1293  & 0.1974 & \\
G5    & 12:43:58.56 & +16:23:01.46  & - & 0.9489 & \\
G6 & 12:43:57.89 & +16:23:01.70 & - & 0.8162 & \\
G7 & 12:43:56.20 & +16:23:06.20 & - & & \\
G8          & 12:43:58.82 & +16:22:57.36 & - & & \\
G9 & 12:43:58.45 &  +16:22:57.12 & - & & \\
G10 & 12:43:57.27 & +16:22:57.82 & 0.286 $\pm$ 0.0890 & 0.2395 & \\
G11      &12:43:56.42 & +16:23:00.71 & - & 0.8877 & \\
G12  & 12:43:56.48 & +16:22:58.21 & - & 0.5586 & 650\\
G13 & 12:43:55.99 & +16:22:59.96 & 0.031 $\pm$ 0.1285 & 0.8888 & \\
G14  & 12:43:58.32 & +16:22:53.00 & 0.624 $\pm$ 0.0390 & 0.6942 & \\
G15 & 12:43:57.76 & +16:22:58.88 & - & 0.5580 & 535\\
G16           & 12:43:59.36    & +16:22:49.21 & 0.466 $\pm$ 0.0562  & & \\
G17 & 12:43:58.46 & +16:22:50.61 & - & & \\
G18 & 12:43:57.84 & +16:22:48.03 & 0.553 $\pm$ 0.0361 & 0.5570 & 342\\
G19 & 12:43:57.31 & +16:22:55.04 & - & & \\
G20 & 12:43:57.23 & +16:22:51.42 & - &  & \\
G21      & 12:43:56.13 & +16:22:54.02 & 0.658 $\pm$ 0.2108 & 0.0956 & \\
G22         & 12:43:56.24 & +16:22:46.73 & 0.571 $\pm$ 0.0412 & 0.5595 & 823\\
G23 & 12:43:57.40 & +16:22:42.94 & 0.113 $\pm$ 0.0730 & \\
G24 & 12:43:58.15 & +16:22:37.84 & 0.126 $\pm$ 0.0423 & 0.1663 & \\
G25 & 12:43:57.89 & +16:22:35.82 &- & 0.5039 & \\
G26 & 12:43:59.00 & +16:22:33.73 & - & & \\
G27 & 12:43:58.78 & +16:22:22.42 & - & & \\
G28 &12:43:57.54 & +16:22:24.81   & 0.470 $\pm$ 0.1262  & 0.5535 & -332\\
G29 & 12:43:56.58 & +16:22:28.29 & 0.457 $\pm$ 0.0232 & 0.4651 & \\
G30 & 12:43:59.47 & +16:22:26.28 & 0.540 $\pm$ 0.0942 & 0.4936 & \\
\hline
\end{tabular}
\hspace{1in}
\label{tab:tab1}
\end{table}

\section{Kinematic Maps of the Giant Nebula} 
\label{sec:section_name}

In order to understand the the main properties of this giant ionized nebula, we fit the most prominent gas emission lines by minimizing $\chi^2$ using the Python package {\tt MPFIT}. After extracting the spectrum in each spaxel, subtracting the continuum and removing the broad emission line, we fit the \oii, \hb\ and \oiii\ lines with a combination of Gaussian profiles, so that all fitted lines share the same central velocity and velocity dispersion.

\subsection{Non-parametric Measurements}  \label{sec:dr}

We performed non-parametric measurements of the emission line profiles following the method described in \citet{Liu2013}, and obtained three maps:

(i) intensity map of the \oiii\  $\lambda$5007 $\mathrm{\AA}$ line.

(ii) $v_{med}$, median velocity map.

(iii) $W_{80}$, line width map. This is defined as the difference between the velocities at 10 and 90 per cent of the cumulative flux: $W_{80} = v_{90} - v_{10}$. For a purely Gaussian velocity profile, $W_{80}$ = 2.563$\sigma$.

\begin{figure*}[htb]
\centering
\gridline{\fig{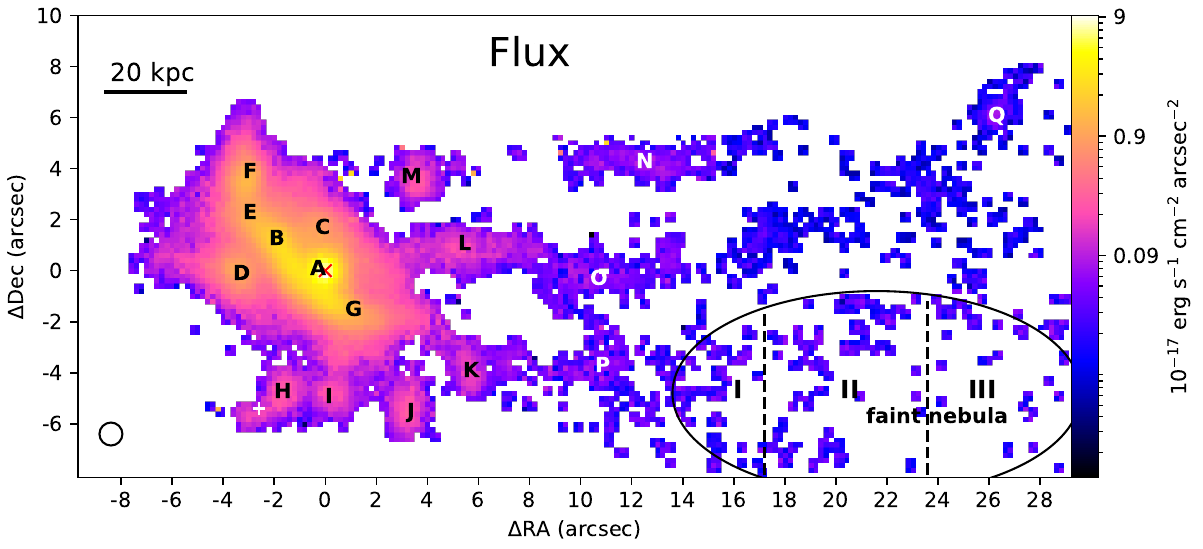}{0.45\textwidth}{(a)}
\fig{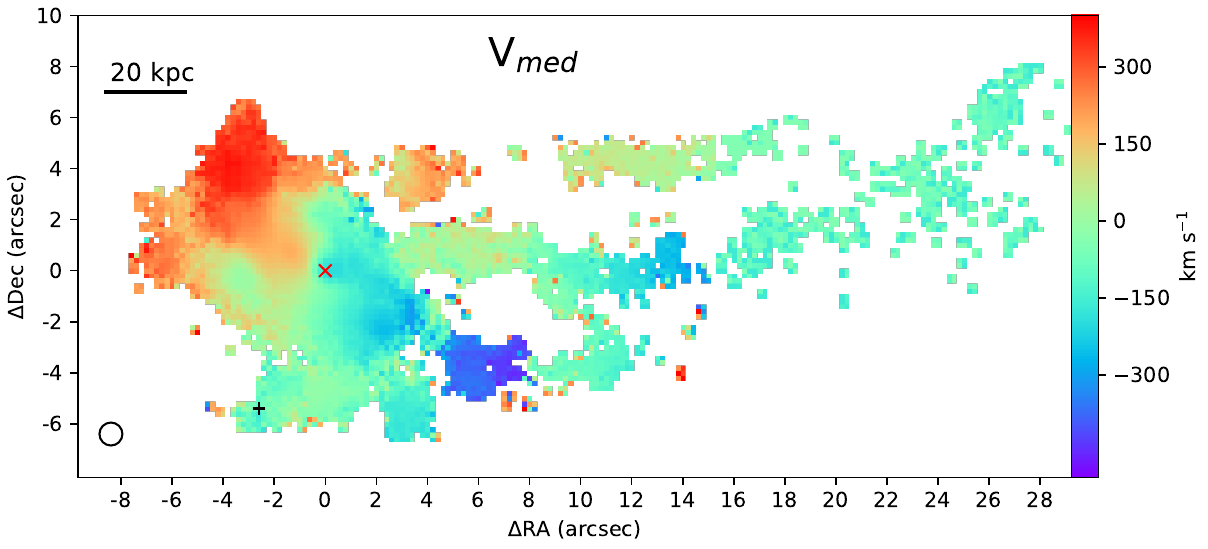}{0.45\textwidth}{(b)}}
\gridline{\fig{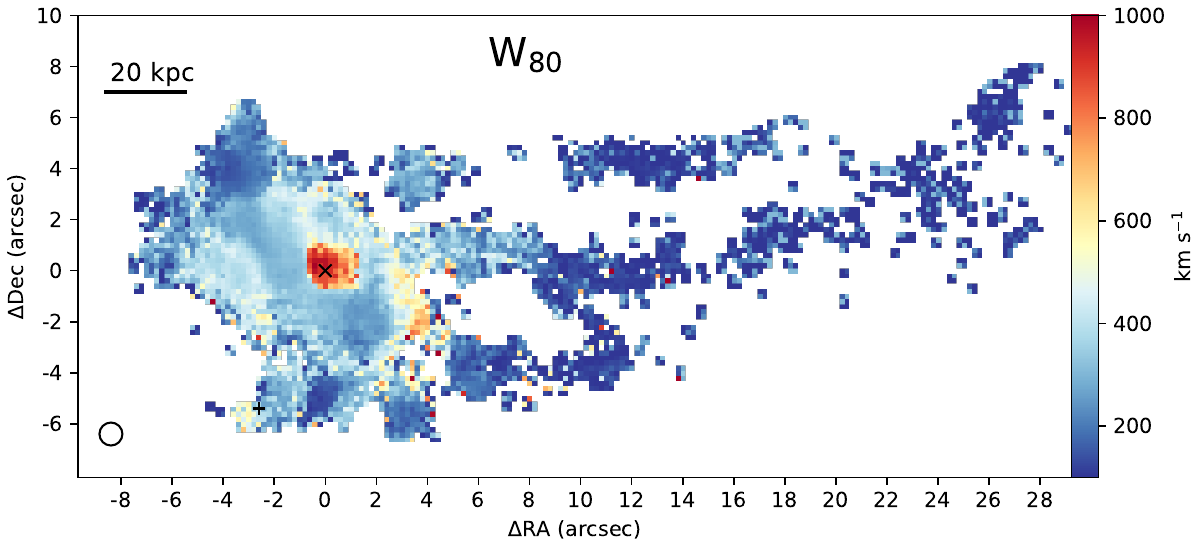}{0.45\textwidth}{(c)}}
\gridline{\fig{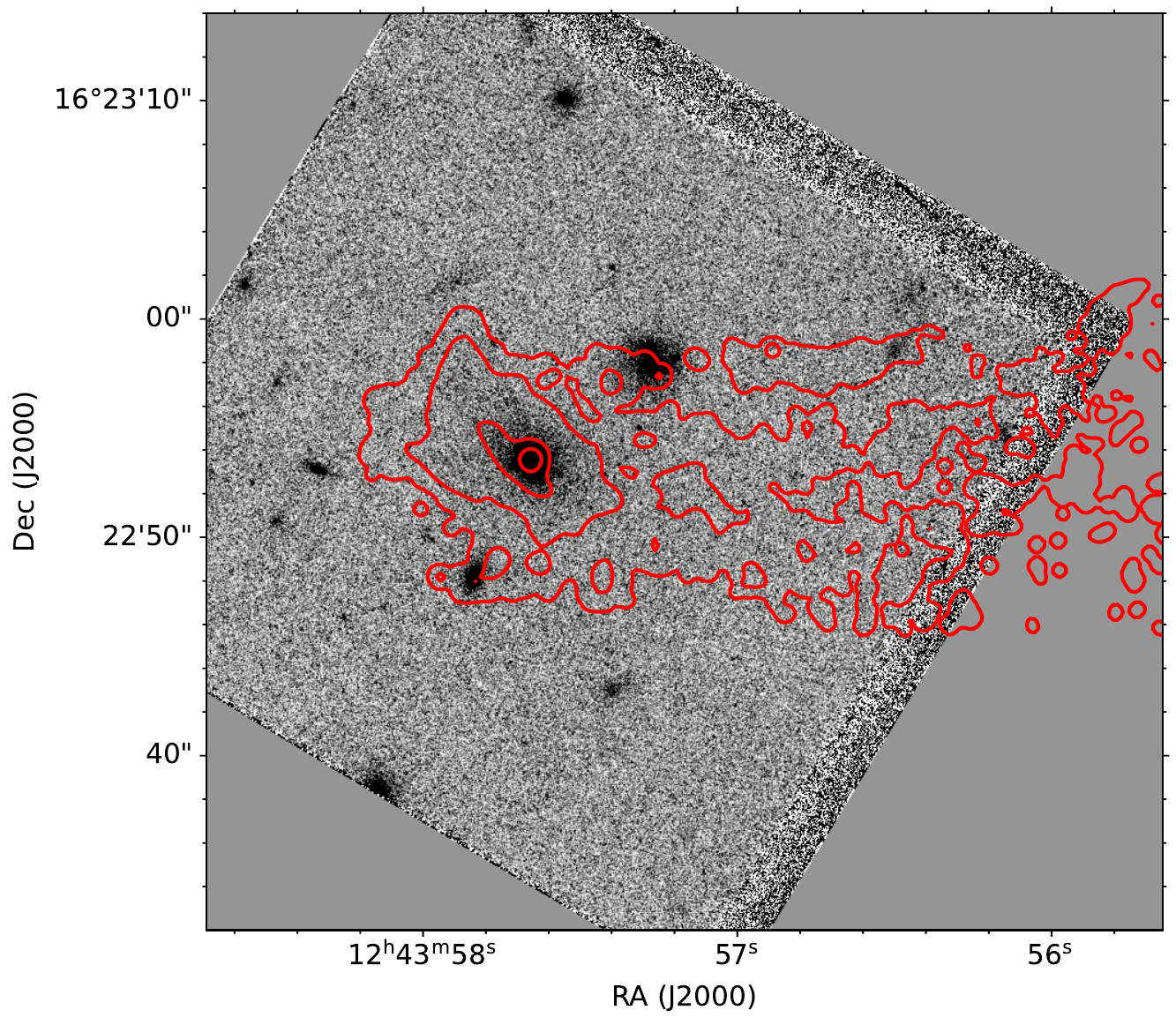}{0.45\textwidth}{(d)}
\fig{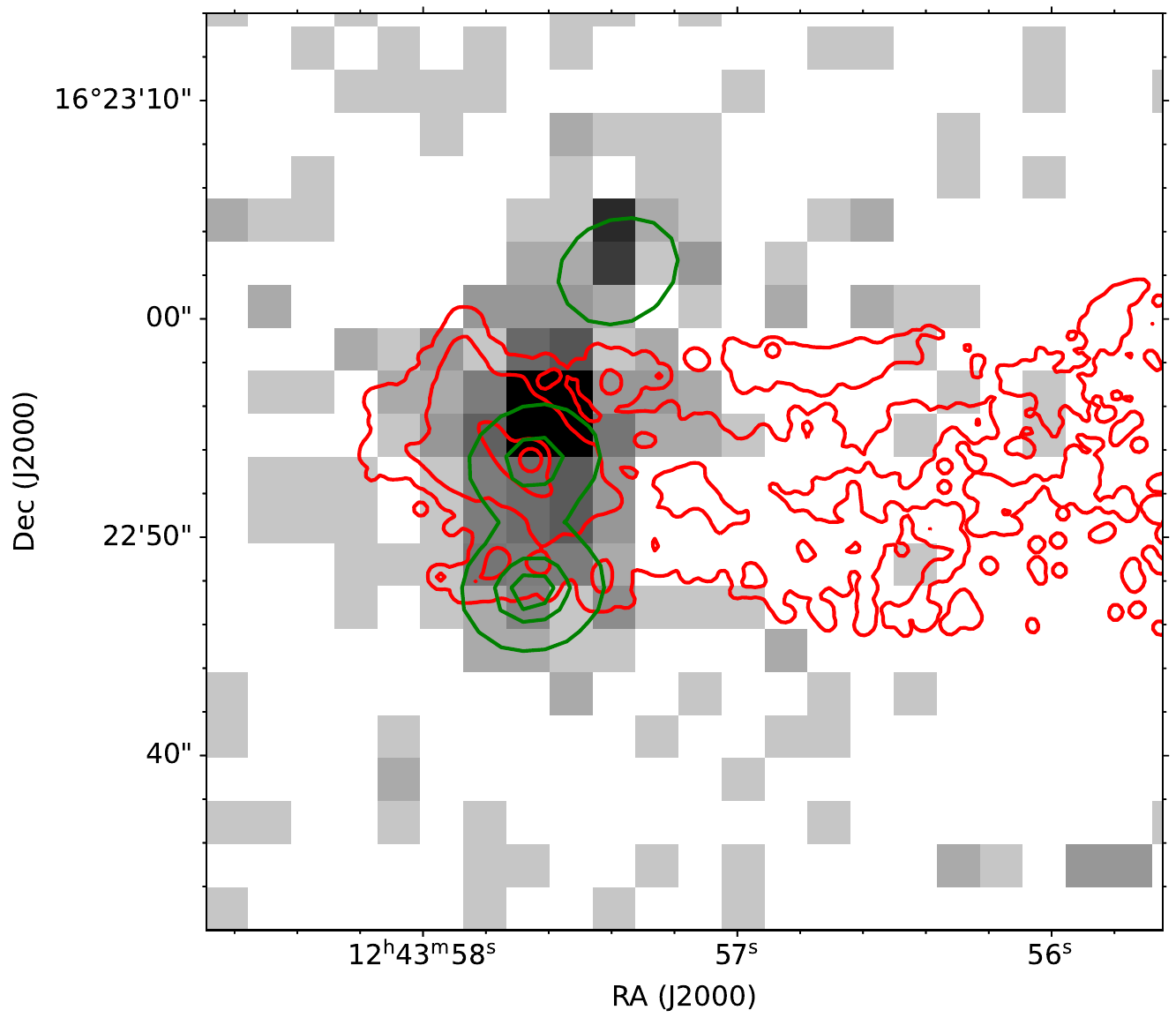}{0.45\textwidth}{(e)}}

\caption{Non-parametric measurements of  \oiii\ nebula. The maps are (a) flux intensity of \oiii\ (erg s$^{-1}$ cm$^{-2}$ arcsec$^{-2}$); (b) median velocity (km s$^{-1}$); (c) line width ($W_{80}$, km s$^{-1}$); (d) \oiii\ surface brightness contours overlaid on the HST WFPC2/F675W image of the field marking surface brightness levels of 5, 150, 1500, and 4000 $\times$ 10$^{-20}$ erg s$^{-1}$ cm$^{-2}$ arcsec$^{-2}$; (e) \oiii\ (red) surface brightness contours and 2.99 GHz radio (green) contours observed by VLASS overlaid on the Chandra/ACIS-S image. The maps were obtained by selecting only those spaxels with a S/N of \oiii\ $\lambda$5007 line equal to or higher than 1.5. The red cross marks the position of the quasar 3C 275.1. The PSF (0.9 $\arcsec$) of MUSE data is depicted by the open circle in the lower left of the panels.  Note that, because of the weak signal, the faint nebula image in the ellipse is generated from the data-cube for \oiii\ in velocity ranges of $\sim$ 300 - 800 km s$^{-1}$ in panel (a). Therefore, the velocity and line width maps (b and c) do not show the nebula in the ellipse. 
\label{fig:fig2}}
\end{figure*}

These non-parametric measurements were obtained using the best-fit profiles. Figure~\ref{fig:fig2} shows an overview of the flux distribution and kinematical properties of the ionized gas, as obtained from the fit of the \oiii\ $\lambda$5007 line. 
The maps were obtained by selecting only those spatial pixels with a signal-to-noise ratio (S/N) equal to or higher than 1.5. 

The resulting morphology and kinematics of the nebula are complex.  The flux distribution (Figure~\ref{fig:fig2}a) shows multiple clumps around the quasar, as well as filamentary structures in the west regions. Based on the morphology, kinematics and locations relative to the quasar, we initially refer to different parts of the nebula as the Host nebula (``A-G"), the Half-ring nebula (``H-K"), the West filaments (``L-Q"), and the faint nebula (ellipse region). The Host nebula surrounds the quasar and extends to a projected radii of about 50 kpc from the quasar with \oiii\ surface brightness levels ranging from 0.1 to 9 $\times$ 10$^{-17}$ erg s$^{-1}$ cm$^{-2}$ arcsec$^{-2}$,  with velocities ranging from -300 to 400 km s$^{-1}$, and distinct rotating disk-like structure (see \S~\ref{sec:dr}). It is attached with two long tails over 100 kpc to the west with \oiii\ surface brightness levels ranging from 0.01 to 0.1 $\times$ 10$^{-17}$ erg s$^{-1}$ cm$^{-2}$ arcsec$^{-2}$.
The flux for the faint nebula in the southwest (in ellipse) is obtained from the data-cube for \oiii\ in velocity ranges of $\sim$ 300 - 800 km s$^{-1}$ rather than fitting \oiii\ due to the weak signal. Therefore,  the velocity and line width maps do not show this nebula.

The large scale velocity structure is rather complex. The velocity distribution (Figure~\ref{fig:fig2}b) of Host nebula is remarkably well organized. One part of the nebula is redshifted, while the other is blueshifted. The velocity map shows evidence for a velocity gradient along the northeast-southwest direction, with velocity ranging between $\sim$ +400 km s$^{-1}$ and $-$300 km s$^{-1}$, possibly associated with a rotating disk.  The velocity of the Half-ring nebula relative to the systemic is approximately zero, except that the ``M" and ``K" regions are redshifted and blue-shifted, respectively.  The velocity of ``K" is different with that of other nearby regions, likely representing an independent system.

The map of $W_{80}$ (Figure~\ref{fig:fig2}c) shows elevated velocity dispersion in the central (``A") region, and the value of $W_{80}$ reaches 1000 km s$^{-1}$, which is comparable to that of known quasar outflows (e.g. \citealt{Zakamska2016}), indicating possible association with an AGN outflow. The other regions have a lower velocity dispersion of $\sim$ 150-400 km s$^{-1}$ at the tail.

\subsection{The Host Nebula}  \label{sec:dr}
\begin{figure*}[htb]
\centering
\gridline{\fig{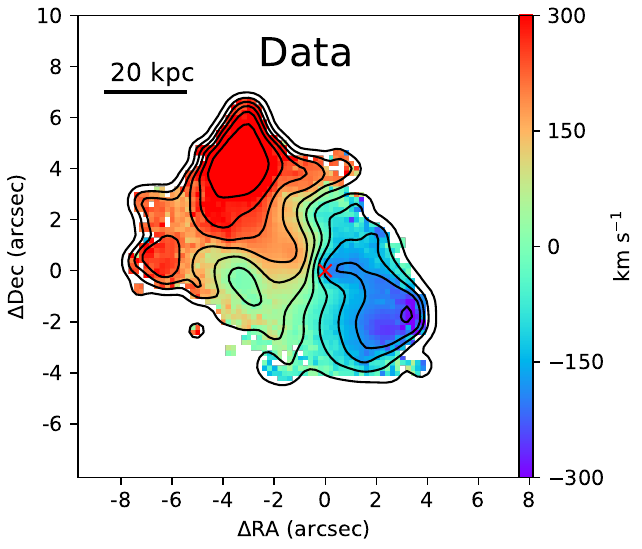}{0.3\textwidth}{(a)}
\fig{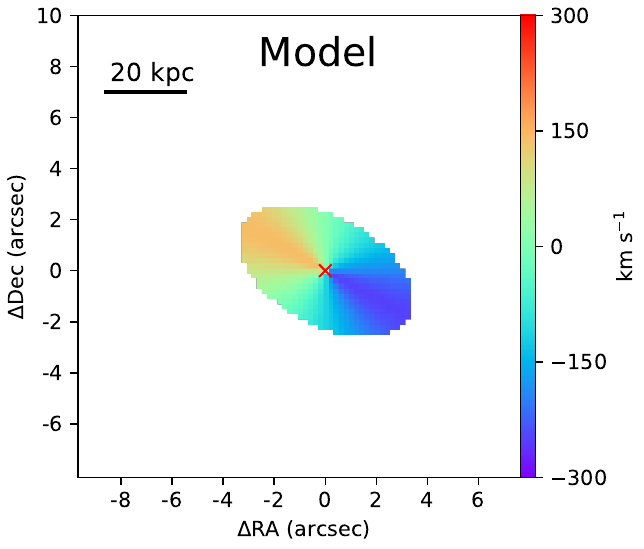}{0.3\textwidth}{(b)}
\fig{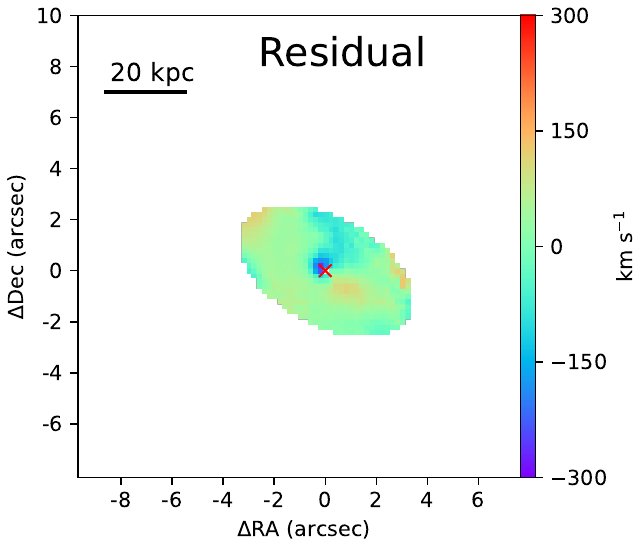}{0.3\textwidth}{(c)}}

\caption{The observed kinematics, the best-fit model using a rotating disk, and the corresponding residual velocity of the Host nebula. Note that the northeast region of the nuclear region is not fitted. The red cross identifies the position of the quasar 3C 275.1.
\label{fig:fig3}}
\end{figure*}

\begin{figure}[h]

\includegraphics[width=0.5\textwidth]{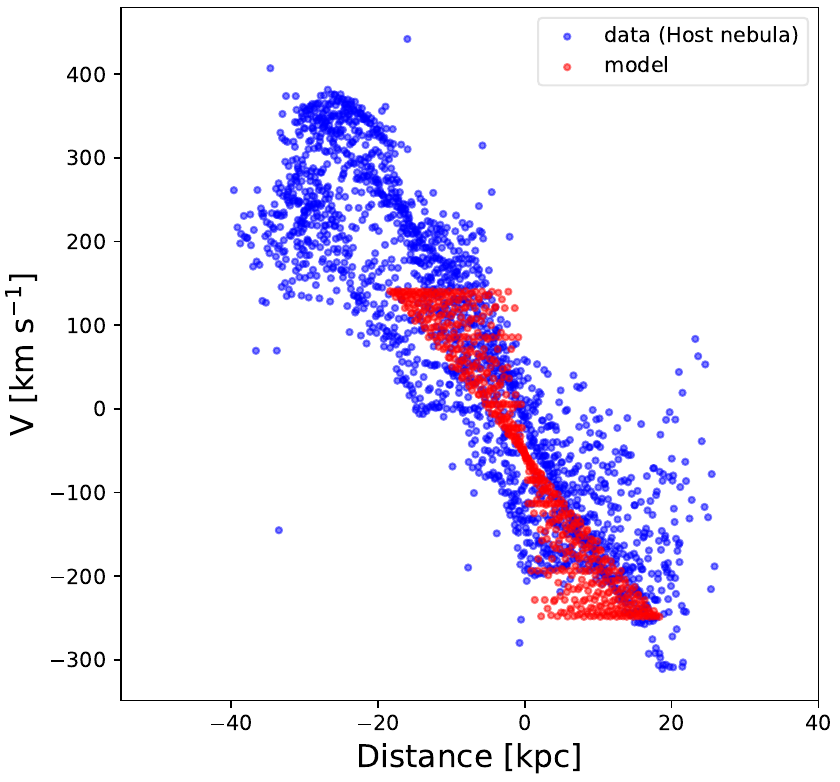}
\caption{ Position-velocity diagram of the model (red) and the observed (blue) taken along the disk major axis (PA = 61$^{\circ}$). 
} 
\label{fig:pv}
\end{figure}

The kinematics of the Host nebula are more complex than the canonical velocity map expected for rotating disks.
We modeled the Host nebula kinematics using a tilted-ring model, and fit with the {\tt BBarolo} \citep{DiTeodoro2015} tool, to test whether the Host nebula kinematics are compatible with a rotation-supported system. This fit has some limitations. Since the Host nebula has more extension in the northeast, and the disk model is symmetric, it cannot be further extended to cover the entire nebula.
The best-fit plots for a rotating disk are shown in Figure~\ref{fig:fig3}. Based on these, we infer an inclination $i = 55^{\circ}$, PA = 61$^{\circ}$, and $v_{rot} \sim 240$ km s$^{-1}$. The best-fit velocity maps still show significant residuals at northeast of the nucleus with $\Delta v \sim 200$ km s$^{-1}$, and $\Delta v \sim -100$ km s$^{-1}$ at nucleus, east and northwest of the nucleus. These features might be associated with a companion (G15) and quasar outflows. We produced position-velocity (PV) diagram, as shown in Figure ~\ref{fig:pv}. In the PV diagram, the range of detected velocities well exceed the projected maximum rotational disk velocity given by the {\tt BBarolo} model. In addition, the velocity amplitude of the Host nebula ($\sim$350  km s$^{-1}$) is higher than that of the normal spiral galaxies ($\sim$200  km s$^{-1}$, \citealt{Vulcani2018}), even the massive spiral galaxies ($\sim$250  km s$^{-1}$), and the size of these nebula is larger than that of normal galaxy disks. Although PV diagram shows solid body rotation, as reported by \citet{Hintzen1986}, the overall kinematics are more complex than this, and the pattern of large velocity residuals in Figure~\ref{fig:fig3} is unlikely to be explained by any simple disk model. It is likely to be a 3D gas distribution with some complex motions.
It implies that additional factors may be causing its kinematics to deviate from a normal rotating disk, such as inflow, outflow, and the interaction with the companion galaxies. 

To examine if the nebula has any relationship with galaxies in the quasar environment, we show \oii\ and \oiii\ emission contours over the HST optical image in Figure~\ref{fig:fig2}d.  The northeast region (``F") is close to galaxy G15, while the eastern (``D") region is close to G9 and G17. There appears to be weak continuum emission in ``D" region (Figure~\ref{fig:fig2}d). The line width in ``F" region is narrower than that from nearby parts of the gas, indicating that the emission line likely arises in its ISM.  The wider line width at  ``D" region is probably the result of the interaction between the companion galaxies and 3C 275.1. 
However, the continuum and the emission lines of G9 and G17 are too faint to obtain spectroscopic redshifts. There may be more dwarf galaxies around the quasar 3C 275.1 with weak continuum, but current imaging data cannot indicate if there are any interaction between these galaxies and the quasar.

\subsection{The Half-ring Nebula}  \label{sec:dr}

3C 275.1 is surrounded by multiple clumps (``H-M"), these clumps are $\sim$ 30-40 kpc away from the quasar. Clumps ``L" and ``M" are more continuously connected to the West filaments and are discussed together next. ``H" region is spatially coincident with companion galaxy G18, and the plus marks the position of G18 (Figure~\ref{fig:fig2}). The surface brightness contours of clump ``H" is somewhat elongated in the southeast–northwest direction, exhibiting an offset (0.5 \arcsec) between its position of surface brightness peak and position of the G18 (Figure~\ref{fig:fig2}d).
Together, the morphology and spatial coincidence with G18 suggest that this elongated clump likely arises from tidal stripping of the ISM of G18 (e.g. \citealt{Decarli2019}). G18 is tidally stretched along its course, thus leading to the spatial extent between the southeast and northwest wings of the G18.
The other half-ring of clumps might be consistent with a stream of tidal debris.

The majority of the entire nebula is dominated by \oiii\ emission, but the ``I" region has greater \oii\ emission (Figure  ~\ref{fig:fig4} and~\ref{fig:fig5}). ``I" also shows a spatial trend in the ionization-state-sensitive O32 and \oiii/\hb\ line ratio maps (Figure~\ref{fig:fig5}), which implies the ionization degree is lower in the south and higher near the quasar. In addition, \oiii\ line profile is different from \hb\, which is wider (Figure~\ref{fig:fig4}). This may be related to the presence of jet, because the projection of the radio lobe is in the ``I" region (Figure~\ref{fig:fig2}e). However, the southward radio jet is expected to be moving away from observer\citep{Liu1990}, while the gas in ``I" region is slightly redshifted or at the systemic velocity. The line width of ``I"  is also narrower than that of other nearby clumps.  Weak continuum emission can be detected after stacking the spectra, but its surface brightness peak is not coincident with any continuum sources in the HST image.  It is plausible that the emission line of ``I" arises from the ISM of the dwarf or young stellar populations in the tidal tail.

The velocity distribution of ``K" shows that it is likely an independent velocity system, more blueshifted than other regions. The  morphology of ``K" is somewhat elongated in an east-west direction. 
As ``K" is spatially adjacent to the galaxy G20, the emission line could be associated with G20. 
This nebula could also arise from RPS of the ISM of a faint dwarf galaxy as it moves through the hot halo of the quasar host group (e.g., \citealt{Chen2019a}). We do not rule out the possibility that ``K", like ``M" and ``L", is associated with the filament to the east and west.

\subsection{The West Filaments}  \label{sec:dr}

Together, the flux distribution and velocity maps all suggest that ``M" -``N", ``L"-``O"-``P"/``Q", and the faint nebula in the ellipse can be regarded as a whole diffuse structure. ``M" and ``L" are probably the footpoints of the long filaments. ``K" may belong here as well. Hence we define these as the West filaments. 

These filamentary structures are long and narrow with a width of $\sim$ 1.5\arcsec (10 kpc). They exhibit a line width of  $W_{80} \sim$ 200 - 400 km s$^{-1}$.
The ``M"-``N" filament extends to $\sim$ 100 kpc west of the quasar and intersects with the Host nebula at $r \sim$20 kpc.  The velocity at ``M" is consistent with the velocity at the Host nebula it connects to, and the velocity at ``N" is close to 0 km s$^{-1}$. The velocity of this filamentary structure (``M" -``N") increases as it gets closer to the quasar. The morphology and kinematics of this filament can be explained by the gases arising from ongoing RPS of ISM (e.g. \citealt{Azevedo2023}) or cool, filamentary accretion (e.g. \citealt{Johnson2022}). However, this velocity gradient does not seem to apply to the ``L"-``O"-``Q" filament, which has an irregular velocity gradient that is inconsistent with a simple case of RPS or filamentary accretion. In addition, the ``L"-``O"-``P" structure shows regular velocity gradient, but the morphology of this structure is neither like accretion inflow nor like ram pressure stripped tails.

\begin{figure}[h]

\includegraphics[width=0.5\textwidth]{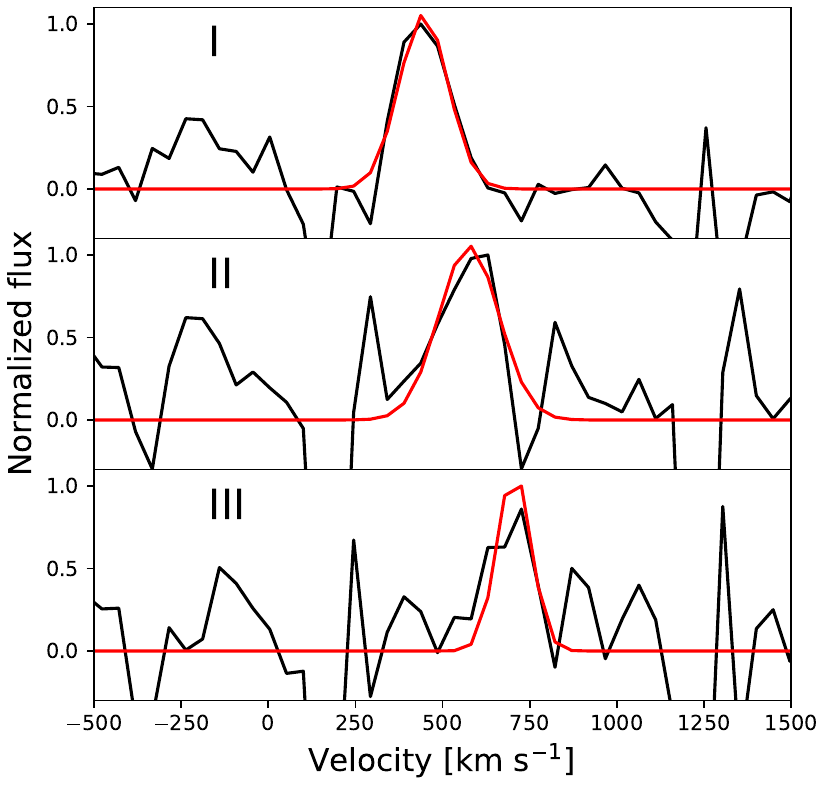}
\caption{ Spectra (black) and best-fit spectral (red) for three regions in faint nebula. Locations of these regions are labelled by their IDs in Figure ~\ref{fig:fig2}). 
} 
\label{fig:fiant}
\end{figure}

\begin{figure*}[htb]

\includegraphics[width=0.9\textwidth]{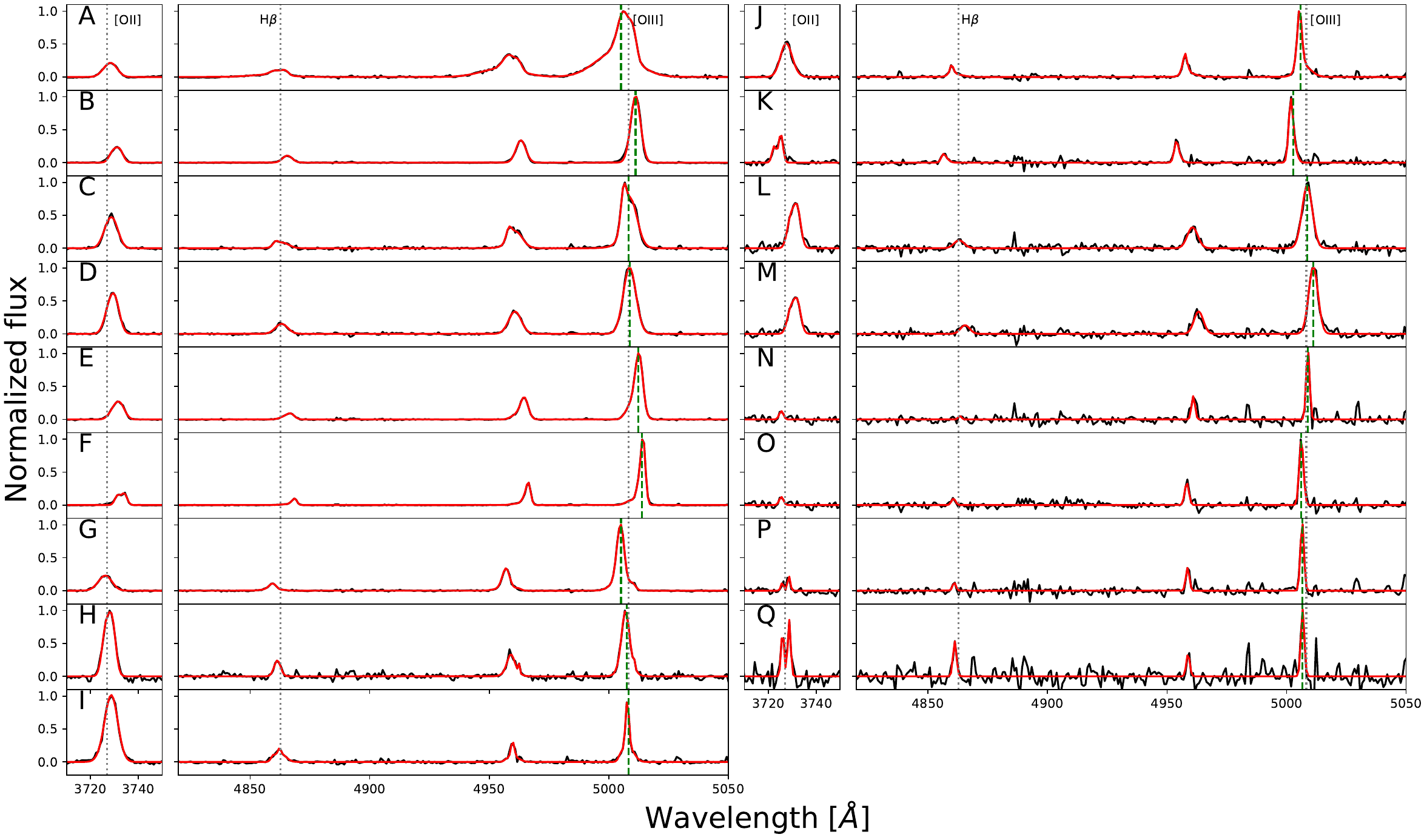}

\caption{ Examples of nebular spectra (stronger lines) and best-fit spectral models for multiple regions, including central (``A''-``G''), and outer regions (``H'' - ``Q''). The median and zero velocity is marked by green dotted and grey dashed lines, respectively. Locations of these regions are labelled by their IDs in Figure ~\ref{fig:fig2}. The extracted spectrum is shown as solid black lines and the best-fit models are shown as red solid lines.
} 
\label{fig:fig4}
\end{figure*}

The faint nebula is redshifted with velocity ranging from 300 to 800 km s$^{-1}$, and the velocity increases with distance (Figure ~\ref{fig:fiant}). 
Such one-sided tail is a signpost RPS feature, and the known jellyfish galaxies with such feature are often observed in galaxy clusters and groups (e.g. \citealt{Bekki2009}). There is no clear velocity gradient along the filaments, as expected for a typical RPS case. 
Given the complexities in the western filaments, we suggest that this system is inconsistent with a simple RPS scenario. It remains plausible that this system is experiencing RPS with some complicating factors that are not yet understood given the current data.

\section{Discussion} 
\label{sec:discussion}

\subsection{Physical Conditions of the Giant Nebula}  \label{sec:dr}

AGN, shocks and young stellar populations all may contribute to ionize the nebula at such a large scale (e.g. \citealt{Liu2013,Epinat2018,Chen2019a}). To better explore the properties of the nebula, we selected several representative regions and extracted their spectra to infer physical conditions and the origin of the ionized gas. 

\begin{figure*}[htb]

\includegraphics[width=0.45\textwidth]{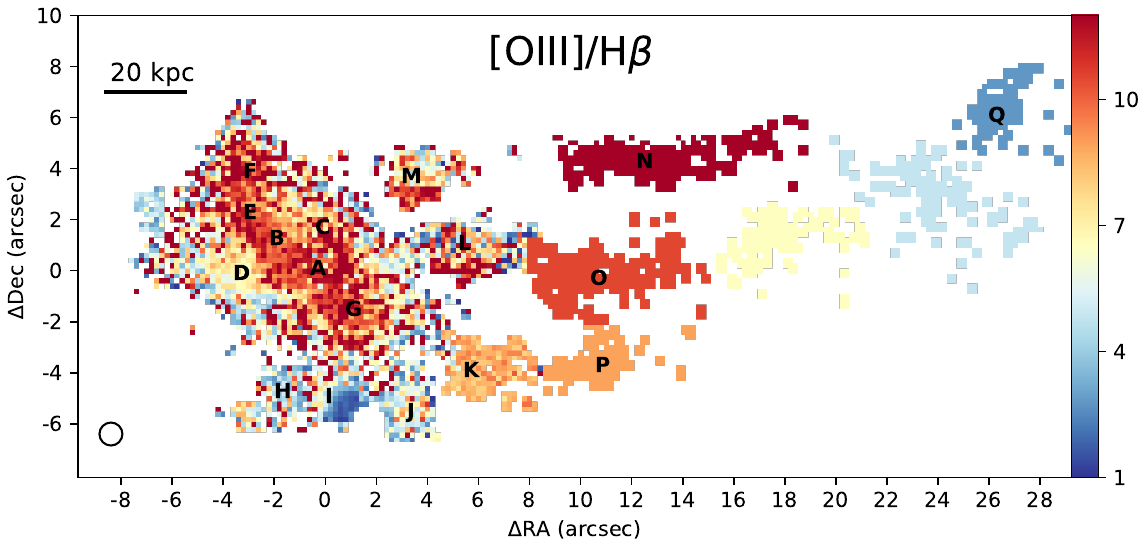}
\includegraphics[width=0.45\textwidth]{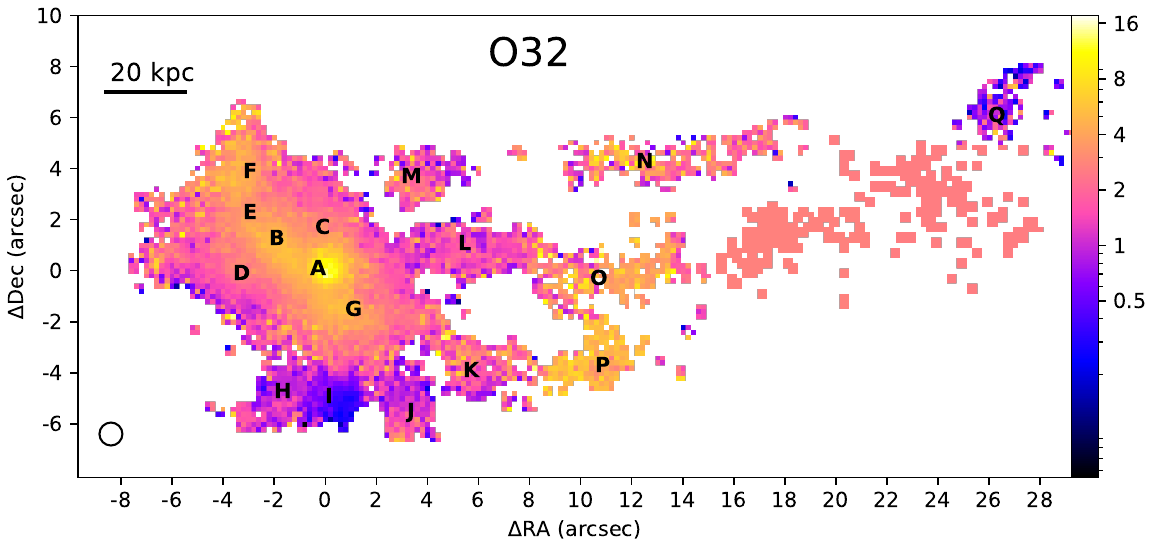}
\caption{The intensity ratio of \oiii\ 5007 to the narrow component of \hb\ image and the line ratio O32 (O32=\oiii$\lambda\lambda$5007+4959/\oii$\lambda\lambda$3727+3726 image). The ratio persists at a constant level ($\sim$10) in the whole field of view, except for the ``I"  and ``Q" where the ratio persists $\sim$ 3. Due to the SNR of the emission line in the outer region, we bin each of the clumps in the outer region (``N-Q") of \oiii/\hb\ map while we bin the region between ``O" and ``Q" in O32 map, and give the average values of these regions.
}
\label{fig:fig5}
\end{figure*}

We measure the the electron density using \oii\ doublet , and quantify the degree of ionization using \oiii/\hb\ and O32 (O32=$\frac{\oiii\lambda5007+\oiii\lambda4959}{\oii\lambda3727+\oii\lambda3726}$) which are ionization state-sensitive strong and weak line ratios . These line ratio measurements and electron density are reported in Table ~\ref{tab:tab2}, and O32, \oiii/\hb\ maps shown in Figure~\ref{fig:fig5}.  We use Python package {\tt Pyneb}, and assume an electron temperature of 10000K, to estimate the electron density. The Host nebula where the \oii\ doublet is resolved, are in the high density state except for ``C" and ``D", while the other nebulas are in a low density limit except for ``H", ``I", ``P" and ``Q".

The inferred electron density in the Host nebula range from $\sim$ 200 to 300 cm$^{-3}$, which is consistent with the electron density of radio galaxies at a few 100 cm$^{-3}$ \citep{Nesvadba2006,Nesvadba2008}. The mean value of electron density is $n_{\rm e} \simeq 100 \rm cm^{-3}$ measured on the integrated spectrum of the tail, which is comparable to the mean value of electron density in the tail of the jellyfish galaxy \citep{Boselli2019}. However, this value is a light-weighted mean, and is thus probably biased towards the brightest and highest density regions. This gas density matches that observed in star forming regions of galaxies at z $\simeq$ 2 \citep{Sanders2016}. 
we can also estimate the typical recombination time of the ionized gas,
\begin{equation}
\tau_\mathrm{rec} = \frac{1}{n_e \times \alpha_A}
\end{equation}
where $\alpha_A$ is the total recombination coefficient ($\alpha_A$ = 4.2 $\times$ 10$^{-13}$ cm$^{3}$ s$^{-1}$, \citealt{Osterbrock2006}). $\tau_\mathrm{rec} \sim 10^{3}$ yr, which is much shorter than the typical dynamical timescale of the tail ($\tau \sim 10^{8}$ yr, e.g. \citealt{Smith2022}). This very short recombination time suggests that, as in local galaxies, a source of gas excitation within the tail must be present. This off-nuclear ionizing source could be young stellar population, which is similar to the observed jellyfish galaxies with extended tails of star forming regions (e.g. \citealt{Poggianti2017}), or it could be shocks or collisional ionization (e.g. \citealt{Fossati2016}), also in line with theoretical expectations \citep{Tonnesen2007}.

The nebula also shows a spatial trend in \oiii/\hb\ and O32 line ratio. The distributions of the \oiii/\hb\ ratio diagnostics are almost constant across the whole nebula ($\sim$ 10), except for the  ``H-J" ($\sim$ 5) and ``Q" ($\sim$ 2) region (see Table~\ref{tab:tab2} and Figure~\ref{fig:fig5}). These values (\oiii/\hb\ $\sim$ 10) indicate that the nebula has a higher degree of ionization than other jellyfish galaxies (\oiii/\hb\ $<$ 3, e.g. \citealt{Poggianti2017,Boselli2019,Bellhouse2019}).
We note that the degree of ionization of ``C" and ``D" are lower than that of regions ``F" -``E" -``B" -``G", which forms an arc-like structure. In addition to AGN photoionization, such an arc-like structure could be heated by shock waves associated with motion in the group, therefore they show higher degree of ionization. The level of ionization is lower in ``I" and ``Q" region, indicating the gas in these region may be ionized by young stellar populations.

Spatial trend of O32 is similar to \oiii/\hb, the higher \oiii/\hb, the larger O32. The arc-like structure has an O32 of around 5, ``A" has a higher O32, which is related to AGN, and  ``C" and ``D" has a lower O32 (see Table~\ref{tab:tab2}).
The majority of the nebula is \oiii\ dominated but the ``I" region has greater \oii\ emission.  Both \oiii/\hb\ and O32 indicate that the Half-ring nebula (``H-K") has lower ionization degree than that of the West filaments (``L-Q") except ``Q", which may be related to electron density and shocks.

\begin{figure}[h]

\includegraphics[width=0.5\textwidth]{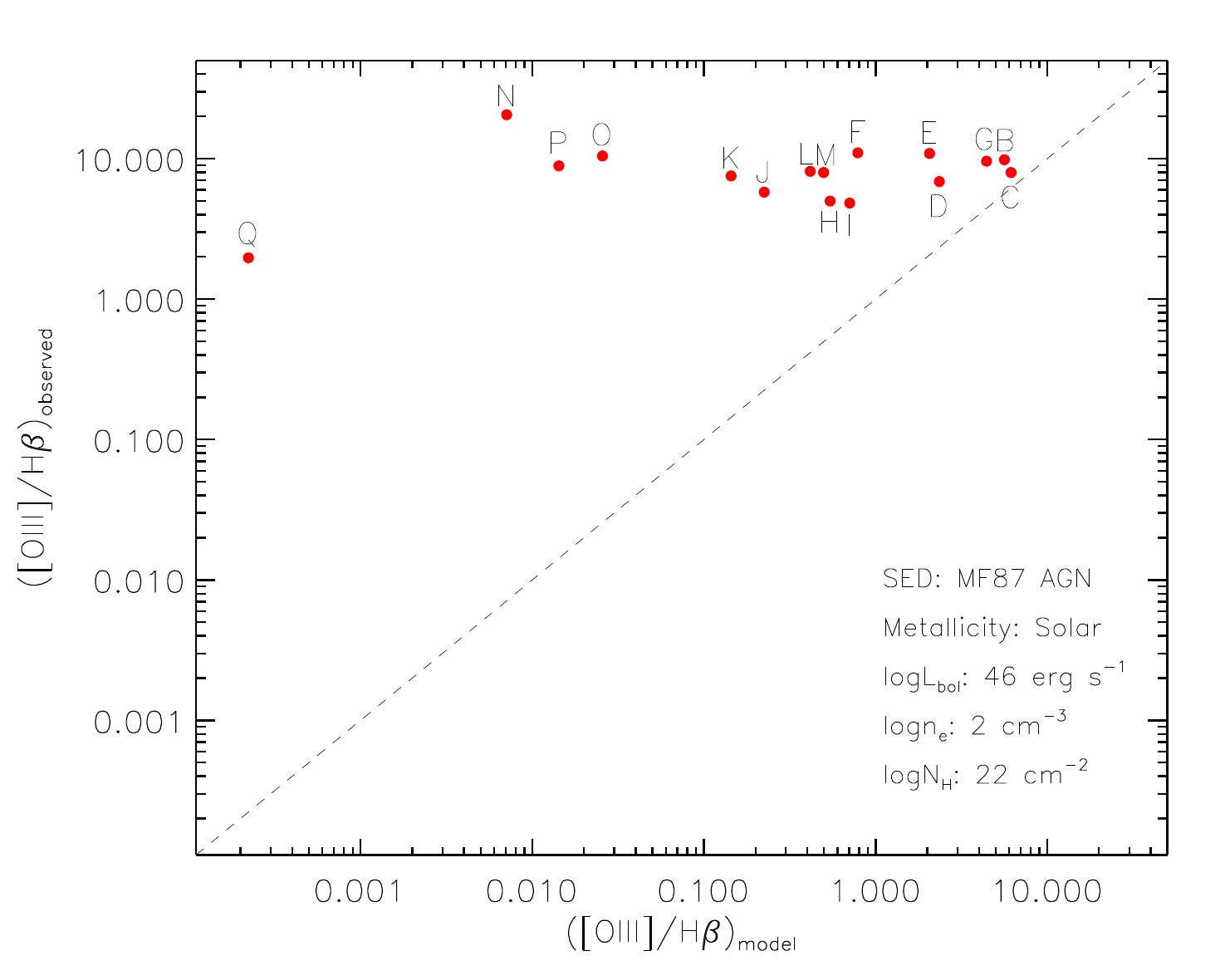}
\caption{ Comparison between the model-predicted \oiii/\hb\ line ratio and the spectrum-observed line ratio in different regions. The dashed line represents a 1:1 relationship.
} 
\label{fig:fig6}
\end{figure}

To better understand its origin, we performed a photoionization simulation assuming that the nebula is illuminated by a central ionizing source with a typical SED of AGN defined by \citet{Mathews1987a} (hereafter MF87). Using the observed quasar continuum at 5100\AA, we have an estimated bolometric luminosity of the ionization source, $L_{\rm bol} \approx 10^{46}$ erg~s$^{-1}$. For simplicity, we assume that the nebula has a slab shape, and a metallicity of solar abundance. The electron density of the nebula is roughly fixed to $n_{\rm e}=100$ cm$^{-3}$, as a typical estimate of the electron density (see Table~\ref{tab:tab2}). The total column density is fixed to a large value of $N_{\rm H}=10^{22}$ cm$^{-2}$, assuming that the nebula is thick enough. The simulation is executed using the {\tt CLOUDY} C17.01 photoionization code \citep{Ferland2017}. We export the model-predicted \oiii/\hb\ line ratio for each nebula region, and compare them with observed line ratio (Figure~\ref{fig:fig6}). It can be seen that the observed line ratio in the whole nebula is higher than its model-predicted values, especially in the tail, possibly because this nebula have additional ionized source besides the quasar, such as young stellar populations or contribution from shocks.

\begin{table*}[htb]\footnotesize
\caption{Summary of emission-line flux,  electron density and line ratio measurements for extracted regions in the nebula}
\hspace{-0.5in}
\begin{tabular}{ccccccc}
\hline
\hline

 Region & Distance & n$_{e}$ &  \oiii/\hb     &  O32 (\oiii/\oii)&   Flux$_{\oiii \lambda5007}$  & Flux$_{\oii \rm doublet}$ \\
 & (arcsec/kpc) & (cm$^{-3}$) & & & $(10^{-17}$ erg s$^{-1}$ cm$^{-2}$ arcsec$^{-2}$) & (10$^{-18}$ erg s$^{-1}$ cm$^{-2}$ arcsec$^{-2}$) \\
\hline
A    & 0  & 213 $\pm$ 237 & 12.67 $\pm$ 0.02 & 11.8 $\pm$ 0.12 & 148.14 $\pm$ 1.21  & 167.32 $\pm$ 20.68                                                    \\    
B    & 2.16/13.94 & 198 $\pm$ 15 & 9.83 $\pm$ 0.01& 5.25 $\pm$ 0.12  &  46.64 $\pm$ 0.06 & 118.41 $\pm$ 14.32  \\
C    & 2.01/12.95 & 26 $\pm$ 13 & 7.97 $\pm$ 0.06 & 2.92 $\pm$ 0.06  & 17.83 $\pm$ 0.14 & 81.53 $\pm$ 4.69 \\
D     & 3.4/21.90 & $<$ 5 & 6.87 $\pm$ 0.03 & 2.33 $\pm$ 0.05 & 18.11 $\pm$ 0.10  &103.53 $\pm$ 4.70  \\
E    & 3.56/22.94 & 400 $\pm$ 21 & 10.89 $\pm$ 0.02 & 3.94 $\pm$ 0.07  & 27.19 $\pm$ 1.75 & 91.99 $\pm$ 0.001 \\
F    & 4.69/30.19 & 325 $\pm$ 26 & 11.01 $\pm$ 0.02 & 4.98 $\pm$ 0.01  & 26.69 $\pm$ 0.22 & 71.42 $\pm$ 0.93  \\
G   & 2.55/16.40 & 339 $\pm$ 33 & 9.60 $\pm$ 0.03 & 3.90 $\pm$ 0.22  & 19.55 $\pm$ 2.27 & 66.86 $\pm$ 13.30   \\
H          & 5.13/33.02 & 165 $\pm$ 51 & 4.99 $\pm$ 0.09 & 1.15 $\pm$ 0.27  & 4.58 $\pm$ 1.14  & 52.96 $\pm$ 9.4 \\
I   & 4.82/31.02    & 91 $\pm$ 20 & 4.83 $\pm$ 0.04 & 0.50 $\pm$ 0.09  & 4.71 $\pm$ 0.12 & 124.55 $\pm$ 11.23\\
J    & 6.28/40.43 & - & 5.78 $\pm$ 0.22 & 1.48 $\pm$ 0.18  & 5.24 $\pm$ 0.36 & 47.21 $\pm$ 7.89 \\
K        & 6.93/44.67 & $<$ 24 & 7.54 $\pm$ 0.12 & 2.32 $\pm$ 0.32 & 3.11 $\pm$ 0.19 &  17.87 $\pm$ 5.6 \\
L    &5.46/35.16 & -   & 8.13 $\pm$ 0.11 & 1.60 $\pm$ 0.22  & 3.60 $\pm$ 0.06 &  30.06 $\pm$ 6.60 \\
M         & 5.23/33.72 & $<$ 127 & 7.97 $\pm$ 0.12 & 2.21 $\pm$ 0.03  & 4.27 $\pm$ 0.07 &  25.72 $\pm$ 0.85  \\
N        & 13.61/87.62 & - & 20.56 $\pm$ 0.73 & 6.29 $\pm$ 0.56  & 1.33 $\pm$ 0.06 & 2.82 $\pm$ 1.58 \\
O       & 10.20/65.72 & - & 10.46 $\pm$ 0.23 & 10.6 $\pm$ 0.22  &1.57 $\pm$ 0.04 & 1.95 $\pm$ 0.44\\
P      & 11.64/74.97 & 182 $\pm$ 142 & 8.89 $\pm$ 0.23 & 5.25 $\pm$ 0.14 &1.35 $\pm$ 0.04 & 3.44 $\pm$ 0.49 \\
Q      &  26.88/173.14 & 219 $\pm$ 185 & 1.97 $\pm$ 0.15  & 1.04 $\pm$ 0.09 & 0.65 $\pm$ 0.05    & 8.30 $\pm$ 0.56 \\                  
\hline
\end{tabular}
\hspace{0.5in}
\label{tab:tab2}
\end{table*}

\subsection{Origin of the Ionized Gas}  \label{sec:dr}

Due to the complex dynamics, morphology and presence of some faint galaxies surrounding 3C 275.1, it is reasonable to expect that this giant nebula is related to its group environment, possibly explained by CGM, IGM or ICM in the presence of interactions.

Firstly, RPS can readily produce the large scale ionized gas seen here, shaping the morphology that resembles a jellyfish galaxy with the striking westwards tails. The tails, which have
a cometary shape with a typical surface brightness of a few times 10$^{-19}$ erg s$^{-1}$ cm$^{-2}$ arcsec$^{-2}$, extend up to $\simeq$ 170 kpc from the nucleus. A natural initial interpretation given its one-sided, linear tail is the result of RPS, and there is at least one reported RPS feature that is nearly as long ($\sim$ 100 kpc; e.g. \citealt{Yagi2007}). 

In addition, the host nebula exhibit arc-like structure (``F" -``E" -``B" -``G"). 
This observed feature is similar to that of the jellyfish galaxies (e.g. \citealt{Bellhouse2017}), in which an arc-like structure is bent into the same direction of their stripped gas trails, possibly suggesting the ram pressure compressed this side of the disk.
Therefore, this arc-like structure would be ionized by shocks, which is consistent with the \oiii/\hb\ line ratio of our {\tt CLOUDY} simulation lower than that of the observation(Figure ~\ref{fig:fig6}).
If 3C 275.1 falls to the cluster center with high velocity, it would undergo extreme RPS (e.g. \citealt{Poggianti2016,Azevedo2023}). 
\citet{KrempecKrygier1998} found that 3C 275.1 is located at the center of the group. Based on the group members that have been identified, 3C 275.1 is likely to suffer from the ram pressure, as redshift of 3C 275.1 differs significantly from the average redshift of these group members, indicating that there is a large velocity difference between 3C 275.1 and its group members. 

It is difficult to conclude if a galaxy is undergoing RPS based solely on its optical morphology, especially since there are other cluster-specific processes that also lead to disturbed morphologies, for example tidal effects and/or harassment (e.g. \citealt{Mayer2006,Chung2007}). The elongated morphology could well be the result of tidal interaction as the group members and surrounding gas fall toward the quasar host. 
The current data is still not sufficient to determine whether 3C 275.1 is located at the center of the group, and future observations including X-rays will be needed.  It is also possible that surrounding galaxies are falling into 3C 275.1, nebula arises from RPS of their ISM as they move through the hot halo of the quasar host group (e.g. \citealt{Boselli2019}). 


Moreover, part of the ionized gas can be attributed to AGN outflow. There is presence of an outflow in the nucleus (``A''), which is indicated by the high velocity dispersion of gas and the blueshifted line asymmetry. However, except for the nuclear region, the velocity dispersion is not high (W$_{80} \lesssim 400$ km s$^{-1}$), which is inconsistent with the AGN-driven outflow with wide opening angles (e.g. \citealt{Liu2013}).

In principle, the ionized gas may also arise from stellar feedback. The interaction between galaxies could increase their star formation and supernova rate {\it in situ}, such as in gas rich RPS tails. However, most regions except for ``A'' exhibit low velocity dispersion, which is inconsistent with high velocity dispersion expected from stellar feedback \citep{Rupke2019}. While ``C-D'', ``H-M'' and ``Q'' regions exhibit lower \oiii/\hb\ ratio, line ratio $\sim 10$ in most regions, implying a high-ionization state in general \citep{Baldwin1981}.  

The kinematics and morphology of the nebula are complex, with rotating structure and striking long tails. We suggest that more processes besides RPS are responsible for the nebula, which may experience tidal stripping and AGN outflow.

\section{Summary}
In this paper, we present discovery of a giant nebula around a radio quasar 3C 275.1 at redshift $z = 0.55522$ using VLT/MUSE observation. The flux distribution shows multiple clumps and the kinematic of the nebula are complex. Based on morphology of the nebula, we separate the large nebula into three substructures including a Host nebula, the Half-ring nebula, and the West filaments. We summarise our results below:

(1) A giant nebula around 3C 275.1 is identified, extending a projected distance up to 170 kpc from the nucleus and resembling a jellyfish galaxy formed via RPS. The nebula emits strongly in \oii, \hb, and \oiii, enabling a good estimation of density and ionization state. 

(2) Through spectroscopic measurements, we confirm that 3C 275.1 is located in a dense environment with identified group members. The southeastern region and the westernmost region are spatially coincident with at least four group member galaxies.

(3) The kinematics and morphology of the nebula are more complex than a simple RPS case. In contrast to other jellyfish galaxies, the nebula of this galaxy has a larger size and higher ionization. The rotation curve of the Host nebula does not flatten in the outer, which is different from those of normal spiral galaxies. It also exhibits an arc-like structure which might be a possible sign of RPS with a higher degree of ionization. This is consistent with our {\tt CLOUDY} simulation, indicating that the arc-like structure needs additional ionizing source besides AGN. Overall, there are various possibilities for the origin of the ionized gas in addition to RPS, such as outflow and tidal stripping during galaxy interactions.




\section{Acknowledgments}

We sincerely thank the anonymous referee for thorough and helpful comments that significantly improved the clarity of our work. We thank Dr. Chong Ge for helpful discussion. We thank Dr. Enrico Di Teodoro for suggestions on fitting the disk with BBarolo. J.W. acknowledges the National Key R\&D Program of China (Grant No. 2023YFA1607904). We acknowledge the NSFC grants 12033004, 12333002, 12221003 and the science research grants from CMS-CSST-2021-A06.
This work is supported by the Fund (KP202105) from the Key Laboratory of Polar Science, MNR, Polar Research Institute of China. Q.Z. acknowledges the support from the China Postdoctoral Science Foundation (2023M732955). This research has made use of the services of the ESO Science Archive Facility. Based on observations collected at the ESO under program 0106.B0564(A). This research used observations made with the NASA/ESA {\em Hubble Space Telescope}, and obtained from the Hubble Legacy Archive, which is a collaboration between the Space Telescope Science Institute (STScI/NASA), the Space Telescope European Coordinating Facility (ST-ECF/ESA) and the Canadian Astronomy Data Centre (CADC/NRC/CSA). 

\facilities{VLT (MUSE), HST (WFPC2)}


\software{astropy \citep{Collaboration2013,Collaboration2018}}



\bibliography{./re.bib}
\bibliographystyle{aasjournal}
\end{document}